# A Comparison of Existing Ethernet Frame Specifications

Muhammad Farooq-i-Azam

*Department of Computer Science,
COMSATS Institute of Information Technology,
Lahore, Pakistan.*
fazam@ciitlahore.edu.pk

**Abstract**
*The Ethernet as we see it today is a result of many stages of evolution. As a result, the Ethernet frame format at the link layer also has seen many changes. One might expect a consistent format of link layer frames as we usually study in text books, in actual practice this is not the case. Formats may vary from one set of specifications to another even on a single local area network due mainly to varying implementations of frame. We monitored representative local area networkscomprising approximately of 1000 hosts in total. The traffic was analyzed using linux based packet sniffer tools – ipgrab, ethereal and tcpdump, to study the Ethernet frame format being used by various types of equipment and vendors. We describe our results in this paper. We also try to clarify ambiguities resulting from a comparison of more common and actual frames to the standards set laid out in IEEE 802.3 specification.*

## 1. Introduction

Overwhelming majority of the local area networks employed all over the world is the Ethernet. Being a widely accepted standard, one would expect of an Ethernet frame to be following a standardized format. However, in actual practice, this is not the case. As discussed in [13], there are four formats of the frame at data link layer on Ethernet type local area networks.

For our analysis, we monitored traffic on representative local area networks comprising of approximately 1000 nodes in total. `Ethereal`, `ipgrab` and `tcpdump` – linux based packet sniffer tools - were used to monitor, capture and analyze the traffic. It was found out there were only following three frame types being used on local area networks in Pakistan:

1- Ethernet II
2- IEEE 802.3 CSMA/CD
3- IEEE 802.3 with SNAP

We could not find a single packet of the fourth frame type i.e. Novell frame format discussed in [13]. This frame format has errors and can only transmit Novell IPX which is rare in Pakistan.

Our analysis showed that more than 97% of the frame types are Ethernet II, 2% are IEEE 802.3 CSMA/CD and only approximately 1% are IEEE 802.3 with SNAP. For example, in a typical capture comprising of 6411 packets consisting of 547648 bytes and with an average packet size of 69 bytes, 6318 frames were Ethernet II, 88 frames were IEEE 802.3 and only 5 frames were IEEE 802.3 with SNAP header. As can be seen in Figure 1, the representaion of frames with SNAP header is almost negligible.

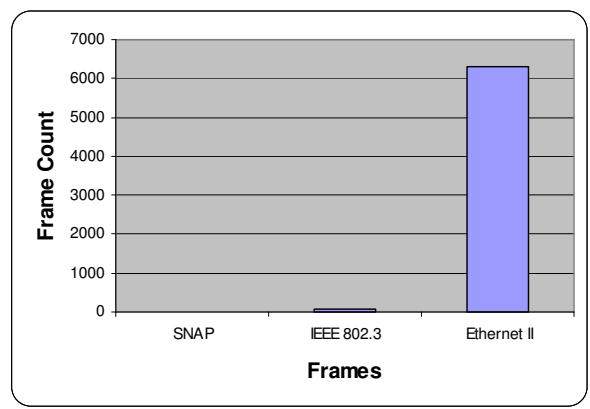

**Figure 1. Population of the Three Frames**

This is contrary to otherwise widespread belief among academicians and the type of frame formats given in reputed texts such as [2] and [14], which are also often used as text books in Pakistan. These texts only describe IEEE 802.3 CSMA/CD frame format and do not mention Ethernet II which constitutes 97% of the traffic on our local area networks. Beneath we describe each of these formats that exist on our local area networks and also givea sample trace from the packets that we captured.

## 2. The Ethernet

Work on an experimental local area network started in Palo Alto Research Center of Xerox Corporation in 1972. By 1976, Bob Metcalf and David Boggs of the research center were able to connect some 100 devices [16].

Digital Equipment Corporation (DEC) and Intel later joined Xerox to formulate the work as a joint specification in 1980. The trio, often termed D-I-X, with each alphabet coming from the first letter of the names of the participating companies, named the specification as the Ethernet.

The same consortium of the three companies released an improved Ethernet specification [3] in 1982. As it was a revised specification, it was termed as Ethernet II by the industry and it widely embraced this newer specification relinquishing its earlier version, which does not exist anymore. It is this version II of the Ethernet that is being widely used on the Ethernet local area networks that we use today.

Format of the frame according to the Ethernet specification II [3] is shown in Figure 2.

| Destination Address | Source Address | Ether Type | Data |
|---|---|---|---|
| 6 bytes | 6 bytes | 2 bytes | upto 1500 bytes |

**Figure 2. Ethernet II frame**

The Ether Type field in this frame specifies the next layer protocol encapsulated in the Data field. In this way, any protocol to be used on top of this frame must have a unique Ether Type. Ether Type values were initially assigned by Xerox, but later as the Ethernet standard was opened up and IEEE based its specification 802.3 on Ethernet II, it took over the responsibility from Xerox in 1997. Any newer protocol to be used above Ethernet II, must be registered with IEEE Registration Authority [6]. Also all existing protocols have a registered EtherType value. For example, Ether Type for IP is 0800 hex and ARP has an Ether Type of 0806 hex.

A trace of a packet using this frame format captured from a live network is shown in Figure 3.

```
Ethernet II
    Destination: ff:ff:ff:ff:ff:ff (Broadcast)
    Source: 00:b0:d0:49:2a:b9 (192.168.1.2)
    Type: ARP (0x0806)
    Trailer: 000000000000000000000000000000000000
Address Resolution Protocol (request)
    Hardware type: Ethernet (0x0001)
    Protocol type: IP (0x0800)
    Hardware size: 6
    Protocol size: 4
    Opcode: request (0x0001)
    Sender MAC address: 00:b0:d0:49:2a:b9
    (192.168.1.2)
    Sender IP address: 192.168.1.2 (192.168.1.2)
    Target MAC address: 00:00:00:00:00:00
    (00:00:00_00:00:00)
    Target IP address: 192.168.1.150
    (192.168.1.150)
```

**Figure 3. Trace of a packet with Ethernet II frame format**

## 3. IEEE 802.3 CSMA/CD

It is common to describe this IEEE specification as the same as the Ethernet, which in fact, is not exactly the case. In fact, after the wide spread industry acceptance of the Ethernet specification, there was a move to standardize the specification by IEEE. The result was IEEE 802.3 CSMA/CD standard [7]. However, this standard deviated from the original Ethernet specification by introducing a 2-byte Length field in place of the Ether Type field. The resulting frame is as shown in Figure 4.

| Destination Address | Source Address | Length | Data |
|---|---|---|---|
| 6 bytes | 6 bytes | 2 bytes | upto 1500 bytes |

**Figure 4. IEEE 802.3 CSMA/CD frame**

Obviously with this frame format, there is no way to determine the next layer protocol. IEEE solved this by an accompanying standard IEEE 802.2 LLC [8]. This specifies a 3 byte header within the Data field after the Length field in the frame shown in Figure 3. Therefore, for it to be used on a link layer, the IEEE 802.3 frame would be as shown in Figure 5.

| Destination Address | Source Address | Length | DSAP | SSAP | CTRL | Data |
|---|---|---|---|---|---|---|
| 6 bytes | 6 bytes | 2 bytes | 1 byte | 1 byte | 1 byte | upto 1497 bytes |

**Figure 5. IEEE 802.3 CSMA/CD with IEEE 802.2 LLC**

The Length field gives length of the frame after itself and can be upto 1500 bytes.

The Control (CTRL) field specifies type of the LLC frame.

The Link Service Access Points (LSAP) i.e. Destination Service Access Point (DSAP) and Source Service Access Point (SSAP) refer to the upper layer protocols/services that the frame uses. In actual practice, both fields are usually set to the same value and serve the same purpose as the Ether Type field in Ethernet II.

However, DSAP and SSAP are only one byte fields whereas Ether Type is a 2-byte field. Furthermore, one bit within DSAP and SSAP is used for some other purposes. So, in effect, there are only 7 bits to specify a next layer protocol which means there can only be $2^7 = 128$ such protocols. On the other hand, Ether Type can specify $2^{16} = 65536$ upper layer protocols.

Furthermore, though there is an LLC protocol type value 6 assigned [19] for the Internet Protocol [12], there is none for the Address Resolution Protocol [10]. IP cannot operate over an Ethernet network without ARP, because without ARP, there would be no mechanism to map a given IP address to an Ethernet address.

It is for these reasons that this frame format could not get acceptability. In fact, link layer services specified by IEEE 802.3 CSMA/CD and IEEE 802.2 LLC were designed to facilitate the use of OSI protocol suite, which failed to get widespread acceptance [13]. Hence, only a small fraction of frames employing this format is spotted on an Ethernet network. And that, perhaps, is due to some network devices using this as their default frame, which can otherwise be configured to use Ethernet II.

Trace of a packet using this frame format is shown in Figure 6.

```
IEEE 802.3 Ethernet
    Destination: 01:80:c2:00:00:00 (Spanning-
    tree-(for-bridges)_00)
    Source: 00:03:31:34:62:c2 (Cisco_34:62:c2)
    Length: 38
    Trailer: 0000000000000000
Logical-Link Control
    DSAP: Spanning Tree BPDU (0x42)
    IG Bit: Individual
    SSAP: Spanning Tree BPDU (0x42)
    CR Bit: Command
    Control field: U, func=UI (0x03)
        000. 00.. = Command: Unnumbered
        Information (0x00)
        .... ..11 = Frame type: Unnumbered frame
        (0x03)
Spanning Tree Protocol
    Protocol Identifier: Spanning Tree Protocol
    (0x0000)
    Protocol Version Identifier: Spanning Tree
    (0)
    BPDU Type: Configuration (0x00)
    BPDU flags: 0x00
        0... .... = Topology Change
                    Acknowledgment: No
        .... ...0 = Topology Change: No
    Root Identifier: 32768 / 00:03:31:34:62:c0
    Root Path Cost: 0
    Bridge Identifier: 32768 / 00:03:31:34:62:c0
    Port identifier: 0x800e
    Message Age: 0
    Max Age: 20
    Hello Time: 2
    Forward Delay: 15
```

**Figure 6. Trace of a packet with IEEE 802.3 CSMA/CD frame format**

DSAP and SSAP i.e. LLC protocol type values are assigned by the IEEE Registration Authority [6].

**3.1 Distinction between Ethernet II and IEEE 802.3 frames**

Now, how do we distinguish an IEEE 802.3 frame from an Ethernet II frame i.e. how would a network interface determine that a particular 2-byte field after the Source Address field is Ether Type or the Length field.

This distinction is made by having Ether Type values larger than the largest value of the Length field which is 1500. In fact, upper layer protocols are assigned Ether Type values equal to or greater than 0600 hex i.e. 1536 decimal [9]. In this way, a value of less than this in the 2-byte field means an IEEE 802.3 frame, whereas a value equal to or greater than 0600 hex implies an Ethernet II frame.

## 4. IEEE 802.3 with SNAP

To address the limitations of IEEE 802.3, an extension to IEEE 802.2 LLC header, called Sub-Network Access Protocol (SNAP) was introduced in [11].

Format of the frame is as shown in Figure 7.

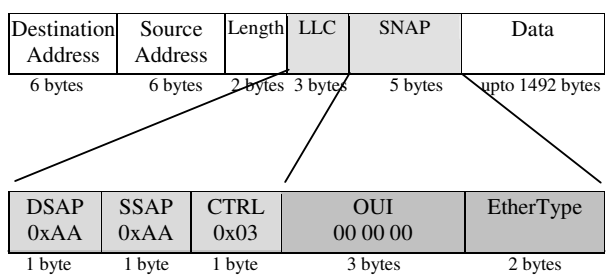

**Figure 7. IEEE 802.3 with IEEE 802.2 LLC/SNAP frame**

As can be seen, this extension, in effect, re-introduces EtherType field to achieve the same function as with Ethernet II, however, with an extra overhead. The encapsulation that can be carried by this frame is only 1492 bytes as compared to 1500 bytes by the Ethernet II frame.

With this frame, DSAP and SSAP have fixed value of AA hex (170 decimal) and the Control field is set to 03. When these values are seen in the LLC header, it is then implied that the subsequent 5 bytes are a SNAP extension.

Within the SNAP extension, first three bytes are reserved for an Organizationally Unique Identifier (OUI). This is the same code as used in the first 3 bytes of an Ethernet address. However, in the case of SNAP, these are always set to 0 as specified in [11]. The EtherType field is the same as in Ethernet II.

So, with this frame format, we, in effect, try to get to the point that we started from i.e. Ethernet II.

Again, we only see this type of frame format only occasionally on an Ethernet local area network.

Trace of a packet with this frame format is shown in Figure 8. It is only this particular type of packets being transmitted by a network device that are using this frame format on the local area network used for the capture. Otherwise, almost all of the packets are Ethernet II.

```
IEEE 802.3 Ethernet
    Destination: 01:00:0c:cc:cc:cc (CDP/VTP)
    Source: 00:03:31:34:73:0b (Cisco_34:73:0b)
    Length: 349
Logical-Link Control
    DSAP: SNAP (0xaa)
    IG Bit: Individual
    SSAP: SNAP (0xaa)
    CR Bit: Command
    Control field: U, func=UI (0x03)
        000. 00.. = Command: Unnumbered
                    Information (0x00)
        .... ..11 = Frame type: Unnumbered frame
                                 (0x03)
    Organization Code: Cisco (0x00000c)
    PID: CDP (0x2000)
Cisco Discovery Protocol
    Version: 2
    TTL: 180 seconds
    Checksum: 0x1787
    Device ID: Switch
        Type: Device ID (0x0001)
        Length: 10
        Device ID: Switch
    Addresses
        Type: Addresses (0x0002)
        Length: 17
        Number of addresses: 1
        IP address: 157.211.1.180
            Protocol type: NLPID
            Protocol length: 1
            Protocol: IP
            Address length: 4
            IP address: 157.211.1.180
    Port ID: FastEthernet0/11
        Type: Port ID (0x0003)
        Length: 20
        Sent through Interface: FastEthernet0/11
    Capabilities
        Type: Capabilities (0x0004)
        Length: 8
        Capabilities: 0x0000000a
            .... ...0 = Not a Router
            .... ..1. = Is  a Transparent Bridge
            .... .0.. = Not a Source Route Bridge
            .... 1... = Is  a Switch
            ...0 .... = Not a Host
            ..0. .... = Not IGMP capable
            .0.. .... = Not a Repeater
    Software Version
        Type: Software version (0x0005)
        Length: 221
```

```
    Software Version: Cisco Internetwork
     Operating System Software
    IOS (tm) C2900XL Software (C2900XL-C3H2S-
     M), Version 12.0(5)XU, RELEASE SOFTWARE
     (fc1)
     Copyright (c) 1986-2000 by cisco
     Systems, Inc.
     Compiled Mon 03-Apr-00 16:37 by swati
  Platform: cisco WS-C2912-XL
    Type: Platform (0x0006)
    Length: 21
    Platform: cisco WS-C2912-XL
  Protocol Hello: Cluster Management
    Type: Protocol Hello (0x0008)
    Length: 36
    OUI: 0x00000C (Cisco)
    Protocol ID: 0x0112 (Cluster Management)
    Cluster Master IP: 0.0.0.0
     UNKNOWN (IP?): 0xFFFFFFFF
                    (255.255.255.255)
    Version?: 0x01
    Sub Version?: 0x01
    Status?: 0x21
    UNKNOWN: 0xFF
    Cluster Commander MAC: 00:00:00:00:00:00
    Switch's MAC: 00:03:31:34:73:00
    UNKNOWN: 0xFF
    Management VLAN: 1
  VTP Management Domain:
    Type: VTP Management Domain (0x0009)
    Length: 4
    VTP Management Domain:
```

**Figure 8. Trace of a packet having IEEE 802.3 CSMA/CD with IEEE 802.2 LLC/SNAP frame format[1]**

## 5. Novell frame format

The days when these frame formats and IEEE standards were being formulated, Novell was a dominant player in the networks. Novell came up with a frame format that it called was in compliance to IEEE 802.3 CSMA/CD whereas it was not.

| Destination Address | Source Address | Length | Data |
|---|---|---|---|
| 6 bytes | 6 bytes | 2 bytes | upto 1500 bytes |

**Figure 9. Novell frame format**

It is really novel in the sense that there is no type field to specify an upper layer protocol. Hence, it can only carry Novell's Internet Packet eXchange (IPX) protocol.

First two bytes in an IPX packet specify checksum. However, if these two bytes are FFFF hex, it signifies that checksum is not being used [19]. By convention, checksum is never used and the two bytes are always set to all 1s. This facilitates network interface drivers to distinguish this frame from an IEEE 802.3 frame with or without SNAP extension.

As IPX alone is almost non-existent, the frame format above also rare. We were not able to get a trace of packet on the local area networks that we analyzed.

## 6. Conclusions

The physical layer specification for all frame types is the same. It is only at the data link layer, where these frame types differ.

It is only the Ethernet II frame format that is being used predominantly on Ethernet local area networks and not the textbook version of IEEE 802.3. Analysis of the captures that we made showed that almost 97% of the frames are Ethernet II, whereas IEEE 802.3 with or without SNAP make up only approximately 3% of the traffic.

Format of destination and source address fields is the same for all frame types.

All frames can co-exist on the same link.

Maximum frame length starting from the Destination Address field to the Data field, both inclusive, is 1514 bytes.

Minimum frame length allowed for the same fields is 60 bytes. If the frame is smaller than this size, a padding of usually all 0s is used to achieve the minimum length.

Preamble, Start Of Frame (SOF) delimiter and Frame Check Sequence (FCS) are used only on the physical layer. Therefore, these are not discussed here.

## 7. References

[1] Borella, Mike, "ipgrab", http://ipgrab.sourceforge.net **

[2] Comer, Douglas E., "Computer Networks And Internets", Prentice Hall, Second Edition, 1999.

---
[1] Sniffer output has been edited to format the contents to fit within the column.